\documentclass[prc,aps,superscriptaddress, English, A4, 10pt]{revtex4-2}
\usepackage[utf8]{inputenc}
\usepackage{graphicx}
\usepackage{xcolor}
\usepackage{babel}
\usepackage{array}
\usepackage{multirow}
\usepackage{amsmath}
\usepackage{amssymb}
\usepackage{graphicx}
\usepackage[unicode=true,pdfusetitle,
 bookmarks=true,bookmarksnumbered=false,bookmarksopen=false,
 breaklinks=false,pdfborder={0 0 0},pdfborderstyle={},backref=false,colorlinks=true]
 {hyperref}
\hypersetup{
 linkcolor=blue, citecolor=cyan}

\makeatletter

\begin{document}
\title{Nonlinear response of flow harmonics in Gubser flow \\ with participant-reaction planes mismatch}

\author{Xiang Ren}
\affiliation{Department of Modern Physics and Anhui Center for Fundamental Sciences in Theoretical Physics, University of Science and Technology of China, Anhui 230026, China}

\author{Jin-Yu Hu}
\affiliation{Department of Modern Physics and Anhui Center for Fundamental Sciences in Theoretical Physics, University of Science and Technology of China, Anhui 230026, China}
\affiliation{School of Science, Huzhou Normal University, Huzhou, Zhejiang 313000, China}

\author{Hao-jie Xu}
\email{haojiexu@zjhu.edu.cn}
\affiliation{School of Science, Huzhou Normal University, Huzhou, Zhejiang 313000, China}
\affiliation{Strong-Coupling Physics International Research Laboratory (SPiRL), Huzhou Normal University, Huzhou, Zhejiang 313000, China}

\author{Shi Pu}
\email{ shipu@ustc.edu.cn}
\affiliation{Department of Modern Physics and Anhui Center for Fundamental Sciences in Theoretical Physics, University of Science and Technology of China, Anhui 230026, China}
\affiliation{Southern Center for Nuclear-Science Theory (SCNT), Institute of Modern Physics, Chinese Academy of Sciences, Huizhou 516000, Guangdong Province, China}

\begin{abstract}
We investigate the nonlinear response of flow harmonics $v_2,v_4$ to initial-state eccentricities $\epsilon_2,\epsilon_4$ within the Gubser-flow framework. By extending the perturbative solutions of Gubser flow, we derive analytic nonlinear response relations connecting the eccentricities $\epsilon_2,\epsilon_4$ to the flow harmonics $v_2,v_4$. Our results reproduce the well-known result  $v_4/v_2^2 \to 1/2$ in large transverse momentum $p_T$ limit. 
Furthermore, we study the effects of a mismatch between the participant and reaction planes. 
We find that the conventional nonlinear response coefficients acquire an additional factor determined by the participant-plane angles, which is often approximated as statistical noise driven by event-by-event fluctuations.%$1$. 
This factor can modify both the strength but even the sign of the effective nonlinear response coefficient, making it sensitive to the initial configuration of the colliding nuclei.
Our study provides new analytical insight into the origin of collective phenomena in relativistic heavy-ion collisions.
\end{abstract}
\maketitle

\section{Introduction}
In relativistic heavy-ion collisions, two heavy nuclei are accelerated to nearly the speed of light and collide with each other. A large amount of energy is deposited in a small spacetime volume, leading to abundant production of quark-antiquark pairs and gluons. The system subsequently undergoes rapid evolution toward local thermal equilibrium and forms a quark-gluon plasma (QGP), which exhibits nearly perfect-fluid behavior~\citep{STAR:2005gfr,PHENIX:2004vcz,BRAHMS:2004adc,ALICE:2008ngc}. The initial entropy density of the QGP is anisotropic in coordinate space due to the off-center nature of the collisions and event-by-event fluctuations. This anisotropy, usually characterized by the eccentricities $\epsilon_{n}$, generates anisotropic pressure gradients that drive anisotropic collective flow. As the QGP expands and cools, quarks and gluons hadronize, and the resulting hadrons are emitted from the freeze-out surface. Consequently, the initial-state spatial anisotropy is converted into a final-state momentum-space anisotropy. The harmonic structure of the single-particle distribution can be characterized by the flow coefficients $v_{n}$ via the Cooper-Frye formula. Therefore, the harmonic coefficients $v_{n}$ are naturally related to the eccentricities $\epsilon_n$.

For a long time, the relationship between the harmonic coefficients $v_{n}$ and the initial eccentricities $\epsilon_{n}$ has attracted significant attention in the relativistic heavy ion collision community, e.g. see Refs.~\citep{CMS:2013wjq,ATLAS:2014ndd,PHENIX:2014uik} and the references therein. This connection provides access to the transport properties of the QGP medium and thereby constrains its effective hydrodynamics. Experimentally, there is strong evidence that $v_{2}$ and $v_{3}$ respond approximately linearly to the corresponding initial eccentricities $\epsilon_{2}$ and $\epsilon_{3}$~\citep{Alver:2010gr,PHOBOS:2006dbo,ATLAS:2014ndd}. For higher-order harmonics such as $v_{4}$ and $v_{5}$, several studies indicate that they are not directly proportional to $\epsilon_{4}$ and $\epsilon_{5}$. Instead, they receive substantial nonlinear contributions generated by the medium response~\citep{Qiu:2011iv,Gardim:2011re}. For example, $v_{4}$ is strongly influenced by $v_{2}^{2}$, while the dominant nonlinear contribution to $v_{5}$ arises from the coupling between $v_{2}$ and $v_{3}$~\citep{ATLAS:2014ndd}.

Conventionally, the nonlinear response coefficient mapping $v_2^2$ to $v_4$, often denoted as $\chi$, has been interpreted primarily as a macroscopic medium response~\cite{Yan:2015jma}. Under this assumption, $\chi$ depends on QGP transport properties, such as shear viscosity, but remains largely independent of the detailed initial geometric profile~\cite{Zhao:2022uhl}. However, recent developments in probing nuclear structure via heavy-ion collisions challenge this clean separation. Recent phenomenological studies have demonstrated that the nonlinear response coefficient is surprisingly sensitive to the initial-state nuclear deformation, particularly the hexadecapole deformation ($\beta_4$)~\citep{Xu:2024bdh, Wang:2024vjf}. This implies that the experimentally extracted $\chi$ encapsulates not only the dynamical medium response but also an intrinsic correlation with the initial spatial configuration. To unravel why the nonlinear response serves as a probe for nuclear deformation, one must look beyond standard numerical averages and rigorously examine the geometric mapping. Specifically, the orientation mismatch between the initial geometric modes is inherently linked to the structural shape of the colliding nuclei. If the nonlinear mapping retains an explicit mathematical dependence on this initial phase mismatch, it would naturally explain the observable sensitivity to $\beta_2$ and $\beta_4$ deformations.

Although substantial progress has been achieved through numerical simulations, the explicit mathematical mapping between the nonlinear flow harmonics and the initial eccentricities has rarely been studied using analytic solutions of relativistic hydrodynamics. As one of the widely used analytic models, Gubser flow~\citep{Gubser:2010ze,Gubser:2010ui} has been employed to investigate properties of the QGP. The original Gubser solution is azimuthally symmetric in the transverse plane, meaning an initial geometric eccentricity cannot be directly incorporated. To address this limitation, Refs.~\citep{Hatta:2014upa,Hatta:2014jva,Hatta:2015era,Hatta:2015hca} introduced perturbations in the transverse plane and derived perturbative solutions around the Gubser background. Within this extended framework, a linear mapping between the initial eccentricities $\epsilon_n$ and the flow harmonics $v_n$ can be established. 

Nevertheless, the nonlinear response of $v_4$ to $\epsilon_2$ and $\epsilon_4$ has not yet been derived in this setup. Motivated by the recent discovery of the $\beta_4$ sensitivity, it is highly desirable to extend the framework of Refs.~\citep{Hatta:2014upa,Hatta:2014jva,Hatta:2015era,Hatta:2015hca} to analytically investigate the nonlinear response. While an ideal Gubser flow model does not fully quantify all transport dissipative effects, it provides a transparent mathematical framework to isolate the geometric factors. Furthermore, in experimental analyses and numerical simulations, the flow harmonic $v_n$ is measured with respect to the $n$-th event plane (often associated with the reaction plane), while the initial eccentricity $\epsilon_n$ is defined with respect to the corresponding participant plane. In general, these planes differ. This geometric phase difference intrinsically alters the extracted nonlinear response, yet its exact mathematical effect has not been systematically exposed within analytic solutions of relativistic hydrodynamics.

To address this gap, in this paper, we extend the perturbative solutions of Gubser flow to analytically derive the nonlinear response of $v_4$ to $\epsilon_2$ and $\epsilon_4$. Furthermore, we explicitly incorporate the orientation mismatch between the initial participant planes and the final reaction planes, demonstrating how this geometric phase difference modifies the extracted nonlinear response coefficients. This paper is organized as follows: In Sec.~\ref{sec:A-Review}, we briefly review the ideal Gubser flow and its perturbative solutions. In Sec.~\ref{sec:EccetricityV2V4}, we introduce a specific transverse perturbation to analytically establish the nonlinear mapping between the flow harmonics ($v_2, v_4$) and the initial eccentricities ($\epsilon_2, \epsilon_4$). In Sec.~\ref{sec:different_planes}, we distinguish the participant planes from the reaction planes, revealing the modified scaling relations induced by their geometric mismatch. Finally, we summarize our findings and discuss their phenomenological implications in Sec.~\ref{sec:Conclusion}.
Throughout this work, we adopt the metric $g_{\mu\nu}=\textrm{diag}\{-1,1,1,1\}$ in the flat Minkowski spacetime, and $u^{\mu}u_{\mu}=-1$ under this metric.

\section{Brief review of Gubser flow with perturbations \label{sec:A-Review}}

In this section, we first briefly review the analytic solution of relativistic hydrodynamics with Gubser flow \citep{Gubser:2010ze, Gubser:2010ui}. We consider an expanding fluid in Minkowski spacetime and adopt the coordinate system with line element $ds^{2} = -d\tau^{2}+dr^{2}+r^{2}d\phi^{2}+\tau^{2}dy^{2}$, where $\tau$ is the proper time, $r$ is the radial coordinate, $\phi$ is the azimuthal angle, and $y$ is the spacetime rapidity.
The basic equations are given by energy-momentum conservation,
\begin{eqnarray}
\nabla_\mu T^{\mu\nu}=0, \label{eq:EM_01}
\end{eqnarray}
where $T^{\mu\nu}$ is the energy-momentum tensor. For simplicity, we neglect charge conservation and consider an ideal fluid. In this case, $T^{\mu\nu}$ takes the form
\begin{equation}
T^{\mu\nu} =(e+P)u^\mu u^\nu +P g^{\mu\nu},
\end{equation}
where $e$, $P$, and $u^\mu$ denote the energy density, pressure, and fluid four-velocity, respectively.  
To obtain an analytic solution of Eq.~(\ref{eq:EM_01}), Refs.~\citep{Gubser:2010ze, Gubser:2010ui} map the coordinates $(\tau,r,\phi,y)$ to a de-Sitter spacetime via a Weyl transformation. Specifically, after Weyl rescaling, the de-Sitter metric is conveniently expressed in the Gubser coordinates $(\rho,\theta,\phi,y)$ as
\begin{eqnarray}
d\hat{s}^{2} & = & \frac{1}{\tau^{2}}ds^{2}=\frac{1}{\tau^{2}}(-d\tau^{2}+dr^{2}+r^{2}d\phi^{2})+dy^{2}\nonumber \\
 & = & -d\rho^{2}+\cosh^{2}\rho(d\theta^{2}+\sin^{2}\theta d\phi^{2})+dy^{2},
\end{eqnarray}
and the transformation between the two coordinate systems is given by
\begin{eqnarray}
\sinh\rho & = & -\frac{L^{2}-\tau^{2}+r^{2}}{2L\tau},\nonumber \\
\tan\theta & = & \frac{2Lr}{L^{2}+\tau^{2}-r^{2}}, \label{eq:Coordinatetransformation}
\end{eqnarray}
where $L$ is the characteristic length scale of the system and corresponds to the  transverse size of the fluid.
To avoid confusion, within this subsection we use a hat to denote physical macroscopic quantities in de-Sitter space after the Weyl transformation. For example, $e$ is the energy density in Minkowski space, whereas $\hat{e}$ denotes the corresponding energy density in de-Sitter space after the Weyl transformation.

Then, Eq.~(\ref{eq:EM_01}) in de-Sitter space after Weyl transformation becomes, 
\begin{eqnarray}
    \hat{u}^{\mu}\hat{\nabla}_{\mu}\hat{e}+(\hat{e}+\hat{p})(\hat{\nabla}\;\hat{u})&=&0, \nonumber\\(\hat{g}^{\nu\alpha}+\hat{u}^{\nu}\hat{u}^{\alpha})\hat{\nabla}_{\mu}P+(\hat{e}+\hat{P})(\hat{u}\;\hat{\nabla})\hat{u}^{\nu}&=&0.
    \label{eq:Gubser_01}
\end{eqnarray}
where the first line represents energy conservation, and the second line is the relativistic Euler equation or the acceleration equation for the fluid velocity.  
Following the Refs.~\citep{Gubser:2010ze, Gubser:2010ui}, a Bjorken-like solution to Eqs.~(\ref{eq:Gubser_01}) is obtained.
In particular, for the ``static'' flow profile $\hat{u}_\mu=(-1,0,0,0)$ with the equation of state $\hat{e}=3\hat{P}$, Eq.~(\ref{eq:Gubser_01}) gives 
$\hat{e}/\hat{e}_0= (\cosh \rho/\cosh \rho_0)^{-8/3}$, with $\hat{e}_0$ being the energy density when $\rho=\rho_0$. One can transform the Navier-Stokes solution (NS) for velocity and energy density back to Minkowski coordinates $(\tau,r,\phi,y)$,
\begin{eqnarray}
u^\mu_{NS} = (u^\tau_{NS},u^r_{NS},0,0), \label{eq:fluid_velocity}
\end{eqnarray}
with 
\begin{eqnarray}
u_{NS}^{\tau} & = & \cosh\left[\tanh^{-1}\left(\frac{2\tau r}{L^{2}+\tau^{2}+r^{2}}\right)\right],\nonumber \\
u_{NS}^{r} & = & \sinh\left[\tanh^{-1}\left(\frac{2\tau r}{L^{2}+\tau^{2}+r^{2}}\right)\right], 
\end{eqnarray}
and 
\begin{eqnarray}
e_{NS} & = & \mathcal{B} T_{NS}^4=\frac{1}{\tau^{4}}\frac{\mathcal{B} C^{4}}{(\cosh\rho)^{8/3}}= \frac{\mathcal{B}C^4}{\tau^{4/3}}\frac{L^{8/3}}{\left[L^{4}+(r^{2}-\tau^{2})^{2}+2L^{2}(r^{2}+\tau^{2})\right]^{4/3}},
\label{eq:energy_density_01}
\end{eqnarray}
where $T_{NS}$ is the temperature, $\mathcal{B}$ is a coefficient determined by the equation of state, and $C$ is a dimensionless parameter fixed by the chosen initial condition. For simplicity, we consider early-time limit $\tau\ll L$ from now on. In this limit, Eq.~(\ref{eq:fluid_velocity}) reduces to 
\begin{eqnarray}
    u^\mu_{NS}=\left(1, \frac{2\tau r}{L^{2}+r^{2}},0,0\right) +\mathcal{O}(\tau^2/L^2). \label{eq:NS_solution_02}
\end{eqnarray}
According to Eq.~(\ref{eq:Coordinatetransformation}), the limit $\tau\ll L$ also implies $\rho\rightarrow - \infty$ with a finite $r$.

We note that although the fluid velocity $u^\mu$ and energy density $e$ depend on $r$, they have no dependence on the azimuthal angle $\theta$. That is, the system remains rotationally symmetric in the transverse plane $(x=r\cos\theta, y=r\sin\theta)$. Consequently, anisotropic harmonic flow coefficients $v_n$ cannot be generated in this setup.
To extend the original Gubser flow, Refs.~\citep{Hatta:2014upa,Hatta:2014jva,Hatta:2015era,Hatta:2015hca} have introduced a perturbative to the solutions in Eqs.~(\ref{eq:fluid_velocity}) and (\ref{eq:energy_density_01}). Then, the temperature and fluid velocity in the de-Sitter space becomes, 
\begin{eqnarray}
\hat{T} & = & \hat{T}_{NS}[1+\delta(\rho)S(\theta,\phi)],\nonumber \\
\hat{u}_{\mu} & = & (-1,\nu_{\theta},\nu_{\phi},0),
\end{eqnarray}
where $\nu_{\theta},\nu_{\phi}$ are functions of $\{\rho,\theta,\phi\}$,
and $\delta(\rho)$ is the fluctuation of the temperature. Inserting the above perturbation to Eqs.~(\ref{eq:Gubser_01}), yields, up to $\mathcal{O}(\partial^2)$,
\begin{eqnarray}
\frac{1}{\tan\theta}\nu_{\theta}+3\cosh^{2}\rho[S(\theta,\phi)\partial_{\rho}\delta(\rho)]+\frac{1}{\sin^{2}\theta}\partial_{\phi}\nu_{\phi}+\partial_{\theta}\nu_{\theta} =0,
\label{eq:diff_01}
\end{eqnarray}
and
\begin{eqnarray}
\delta(\rho)\partial_{\theta}S(\theta,\phi)-\frac{2}{3}\nu_{\theta}\tanh\rho+\partial_{\rho}\nu_{\theta}&=&0,\nonumber\\
\delta(\rho)\partial_{\phi}S(\theta,\phi)-\frac{2}{3}\nu_{\phi}\tanh\rho+\partial_{\rho}\nu_{\phi} &=& 0. \label{eq:diff_perturbative_01}
\end{eqnarray}
From Eqs.~(\ref{eq:diff_perturbative_01}), we find,
\begin{eqnarray}
\nu_{\phi}(\rho,\theta,\phi) & = & \nu_{s}(\rho)\partial_{\phi}S(\theta,\phi),\nonumber \\
\nu_{\theta}(\rho,\theta,\phi) & = & \nu_{s}(\rho)\partial_{\theta}S(\theta,\phi).\label{eq:General_solution_nu}
\end{eqnarray}
Inserting the relation (\ref{eq:General_solution_nu}) back to Eqs.~(\ref{eq:diff_perturbative_01}), yields, 
\begin{eqnarray}
\partial_{\rho}\nu_{s}(\rho) & = & -\left[\delta(\rho)+\frac{2}{3}\nu_{s}(\rho)\right],
\end{eqnarray}
and Eq.~(\ref{eq:diff_01}) reduces to,
\begin{eqnarray}
    \partial_{\rho}\delta(\rho) & = & 0+\mathcal{O}(e^{2\rho}).
\end{eqnarray}
in the early time limit, i.e. $\tau/L\ll1$ or $\rho\rightarrow-\infty$. 
We find,
\begin{eqnarray}
\delta & = & \delta_0+\mathcal{O}(e^{2\rho}),\nonumber \\
\nu_{s} & = & -\frac{3}{2}\delta_0.\label{eq:perturbation_parameters}
\end{eqnarray}
Here, $\delta_0$ is an arbitrary constant. We have checked that $\delta_0$ cancels after normalization when deriving the eccentricity. Consequently, the final results do not depend on the specific value of $\delta_0$.
For simplicity, we choose $\delta_0=1$. Finally, we transform the solutions back to Minkowski spacetime and find,
\begin{eqnarray}
T & = & T_{NS}[1+ S(\theta,\phi)],\nonumber \\
u_\mu &=& \left(1, u_r^{NS}+\delta u_r, \delta u_\phi, 0\right), \label{eq:v_under_perturbation}
\end{eqnarray}
with
\begin{eqnarray}
    \delta u_{r}&=&-\frac{3\tau L}{L^{2}+r^{2}}\partial_{\theta}S(\theta,\phi),\nonumber\\\delta u_{\phi}&=&-\frac{3}{2}\tau\partial_{\phi}S(\theta,\phi).
\end{eqnarray}
Note that $S(\theta,\phi)$ can be specified according to the chosen physical setup or the imposed initial conditions.

\section{Eccentricity and $v_2,v_4$ with perturbative solutions} \label{sec:EccetricityV2V4}

\subsection{Eccentricity from perturbative solutions}\label{subsec:Eccetricity}

As discussed above, the perturbative solution in Eqs.~(\ref{eq:v_under_perturbation}) still leaves $S(\theta,\phi)$ unspecified. In Refs.~\citep{Hatta:2014upa,Hatta:2014jva,Hatta:2015era,Hatta:2015hca}, $S(\theta,\phi)$ is taken to be proportional to the initial eccentricity $\epsilon_n$, and using the Cooper-Frye formula it was found that $v_n \propto \epsilon_n$. In the present study, we follow the same strategy as in Refs.~\citep{Hatta:2014upa,Hatta:2014jva,Hatta:2015era,Hatta:2015hca} and extend the discussion to the nonlinear response of $v_4$ to the initial eccentricities $\epsilon_2$ and $\epsilon_4$.

We first specify $S(\theta,\phi)$ as
\begin{eqnarray}
S(\theta,\phi) & = & a_{2}\left(\frac{2rL}{L^{2}+r^{2}}\right)^{2}\cos2\phi+a_{4}\left(\frac{2rL}{L^{2}+r^{2}}\right)^{4}\cos4\phi. 
\label{eq:S_theta_phi}
\end{eqnarray}
where $a_{2}$ and $a_{4}$ are constant determined by initial conditions. Below, we will show that $a_{2}$ and $a_{4}$ are related to the eccentricities $\epsilon_{2}$ and $\epsilon_{4}$. 
Before presenting the explicit calculations, we introduce the power-counting scheme used in this work. We assume the following hierarchy,
$1\gg a_2>a_4\gg a_2^2 \gtrsim L\partial \gg a_2 a_4 >a_4^2 \gtrsim L^2\partial^2$.
Here, $L\partial$ plays a role analogous to the Knudsen number, and our calculation can be understood as a gradient expansion. Accordingly, for simplicity we truncate the subsequent calculations at $\mathcal{O}(a_2a_4)$. After relating the parameters $a_2,a_4$ to the initial eccentricities, we will briefly discuss the physical interpretation of the above power-counting scheme.
With this choice of $S(\theta,\phi)$, inserting it into the perturbed solutions in Eqs.~(\ref{eq:v_under_perturbation}) yields the velocity,
\begin{eqnarray}
    u_{r} & = & \frac{2\tau r}{L^{2}+r^{2}}-a_{2}\frac{12\tau L^{2}r(L^{2}-r^{2})}{(L^{2}+r^{2})^{3}}\cos2\phi -a_{4}\frac{96\tau L^{4}r^{3}(L^{2}-r^{2})}{(L^{2}+r^{2})^{5}}\cos4\phi, \nonumber \\
u_{\phi} & = & 3\tau a_{2}\left(\frac{2rL}{L^{2}+r^{2}}\right)^{2}\sin 2\phi+6\tau a_{4}\left(\frac{2rL}{L^{2}+r^{2}}\right)^{4}\sin 4\phi.\label{eq:perturbed_solution_02}
\end{eqnarray}
For convenience, we also derive $T^3$ and $e^{3/4}$ for the specified $S(\theta,\phi)$, 
\begin{eqnarray}
    T^3= e^{3/4}/\mathcal{B}^{3/4}&=&\frac{1}{\tau}\frac{4C^{3}L^{2}}{(L^{2}+r^{2})^{2}}\left[1+\frac{3}{2}a_{2}^{2}\left(\frac{2rL}{L^{2}+r^{2}}\right)^{4}+3a_{2}\left(\frac{2rL}{L^{2}+r^{2}}\right)^{2}\cos(2\phi) \right. \nonumber \\
    & & \left.+\left(3a_{4}+\frac{3}{2}a_{2}^{2}\right)\left(\frac{2rL}{L^{2}+r^{2}}\right)^{4}\cos(4\phi)\right] + \mathcal{O}(a_2a_4). \label{eq:T3}
\end{eqnarray}
Since now the entropy density is $4\mathcal{B}\;T^3/3$, the introduced perturbations generate an azimuthal-angle dependence of the entropy density, leading to a geometric anisotropy. This anisotropy can be characterized by the eccentricity $\epsilon_{n}$, which is a dimensionless parameter that ranges from $0$ to $1$. Larger values of $\epsilon_{n}$ correspond to a stronger deviation from azimuthal symmetry. Following Refs.~\citep{Hatta:2014upa,Hatta:2014jva,Hatta:2015era,Hatta:2015hca}, the eccentricity $\epsilon_{n}$
is defined as, 
\begin{eqnarray}
\epsilon_{n} & = & -\frac{\int d^{2}r\; e^{3/4}\frac{r^{n}}{(L^{2}+r^{2})^{n-1}}\cos n\phi}{\int d^{2}r\; e^{3/4}\frac{r^{n}}{(L^{2}+r^{2})^{n-1}}},
\end{eqnarray}
Here we introduce the weight $\frac{r^{n}}{(L^{2}+r^{2})^{n-1}}$ as a cutoff to avoid divergent integrals. By inserting Eq.~(\ref{eq:T3}) into the definition of $\epsilon_{n}$, we obtain the relation between the parameters $a_{2}$, $a_{4}$ and the eccentricities $\epsilon_{2}$ and $\epsilon_{4}$,
\begin{eqnarray}
\epsilon_{2} & = & -a_{2}+\mathcal{O}(a_2a_4),\nonumber \\
\epsilon_{4} & = & -\frac{36}{35}a_{4}-\frac{18}{35}a_{2}^{2}+\mathcal{O}(a_2a_4).
\end{eqnarray}
Alternatively, we can express $a_2$ and $a_4$ in terms of $\epsilon_2$ and $\epsilon_4$,
\begin{eqnarray}
a_{2} & = & -\epsilon_{2}+\mathcal{O}(a_2a_4),\nonumber \\
a_{4} & = & -\frac{35}{36}\epsilon_{4}-\frac{1}{2}\epsilon_{2}^{2}+\mathcal{O}(a_2a_4).\label{eq:Relation_eccentricity}
\end{eqnarray}
We now see that $a_2$ plays the role of $\epsilon_2$, while truncating the power counting at $\mathcal{O}(a_2a_4)$ is equivalent to keeping harmonic flow coefficients up to $\mathcal{O}(\epsilon_2^2)$.

\subsection{Computing $v_{2},v_4$ from Cooper-Frye formula \label{sec:Cooper-Frye}}

Now, we can derive the anisotropic flow using the Cooper-Frye
formula,
\begin{eqnarray}
(2\pi)^{3}\frac{dN}{dYp_{T}dp_{T}d\phi_{p}} & = & -\int_{\Sigma}p^{\mu}d\sigma_{\mu}f(p^{\mu}u_{\mu}/T_{f})\propto1+2\sum_nv_{n}\cos n \phi_{p}.\label{eq:Cooper_Frye_formula}
\end{eqnarray}
where $T_{f}$ is the freeze-out temperature, $\phi_p$ is the particle azimuthal angle, $f(p)$ is the distribution function, and $\Sigma^\mu$ denotes the freeze-out hypersurface, taken here to be the isothermal surface defined by $T=T_{f}$.
For convenience, we parametrize the freeze-out temperature as
\begin{eqnarray}
    T_f = \frac{CB}{2L}, \label{eq:T_f}
\end{eqnarray}
where $B$ is a dimensionless constant. Since the temperature is a function of the proper time $\tau$, we determine the freeze-out proper time from the condition $T=T_f$,
\begin{eqnarray}
\tau_{f} & = &	\tau_{0}+\delta\tau_{0}a_{2}^{2}+\delta\tau_{2}a_{2}\cos2\phi+\delta\tau_{4(1)}a_{4}\cos4\phi+\delta\tau_{4(2)}a_{2}^{2}\cos4\phi.\label{eq:Freezeout_time}
\end{eqnarray}
where 
\begin{eqnarray}
\tau_{0} & = &	\frac{(2L)^{5}}{B^{3}(L^{2}+r^{2})^{2}},\nonumber\\
\delta\tau_{0}	& = &	\frac{(2L)^{5}}{B^{3}(L^{2}+r^{2})^{2}}\times\frac{3}{2}\left(\frac{2rL}{L^{2}+r^{2}}\right)^{4},\nonumber\\
\delta\tau_{2}	& = &	\frac{(2L)^{5}}{B^{3}(L^{2}+r^{2})^{2}}\times3\left(\frac{2rL}{L^{2}+r^{2}}\right)^{2},\nonumber\\
\delta\tau_{4(1)}	& = &	\frac{(2L)^{5}}{B^{3}(L^{2}+r^{2})^{2}}\times3\left(\frac{2rL}{L^{2}+r^{2}}\right)^{4}.\nonumber\\
\delta\tau_{4(2)}	& = &	\frac{(2L)^{5}}{B^{3}(L^{2}+r^{2})^{2}}\times\frac{3}{2}\left(\frac{2rL}{L^{2}+r^{2}}\right)^{4}.
\end{eqnarray}
Inserting the $\tau_f$ into Eqs.~(\ref{eq:perturbed_solution_02}), we derive the fluid velocity at the freeze-out hypersurface, 
\begin{eqnarray}
u_{\phi} & = & \delta u_{\phi2}a_{2}\sin2\phi+\delta u_{\phi4(1)}a_{4}\sin4\phi+\delta u_{\phi4(2)}a_{2}^{2}\sin4\phi,\nonumber\\
u_{r} & = & u_{r0}+\delta u_{r0}a_{2}^{2}+\delta u_{r2}a_{2}\cos2\phi+\delta u_{r4(1)}a_{4}\cos4\phi+\delta u_{r4(2)}a_{2}^{2}\cos4\phi.\label{eq:velocity_freezeout}
\end{eqnarray}
The expression for the $\delta u_{\phi2},\delta u_{\phi4(1)},u_{\phi4(2)},u_{r0},\delta u_{r0} ,\delta u_{r2},\delta u_{r4(1)},\delta u_{r4(2)}$
are shown in Eqs.~(\ref{eq:delta_u_01}) in Append.~\ref{sec:appendix}.

Next, let us compute the distribution function. For convenience, we define the transverse mass as $m_{T}=\sqrt{m_0^{2}+p_{T}^{2}}$, where $m_0$ is the particle mass and $p_T$ is the transverse momentum. We also introduce the momentum rapidity $Y$, to be distinguished from the spacetime rapidity $y$. Then, $p^{\mu}u_{\mu}$ is given by,
\begin{eqnarray}
p^{\mu}u_{\mu} & = & -m_{T}\cosh(y-Y)+p_{T}u_{r}\cos(\phi-\phi_{p})-\frac{p_{T}u_{\phi}}{r}\sin (\phi-\phi_{p}).
\end{eqnarray}
Similarly, we can express $p^{\mu}d\sigma_{\mu}$ as,
\begin{eqnarray}
-p^{\mu}d\sigma_{\mu} & = & r\tau\left(m_{T}\cosh(y-Y)-p_{T}\cos(\phi-\phi_{p})\frac{\partial\tau}{\partial r}+\frac{p_{T}}{r}\sin (\phi-\phi_{p})\frac{\partial\tau}{\partial\phi}\right)dydrd\phi.
\end{eqnarray}
Eq.~(\ref{eq:Cooper_Frye_formula}) becomes,
\begin{eqnarray}
    (2\pi)^{3}\frac{dN}{dYp_{T}dp_{T}d\phi_{p}}&=&2\int drd\phi\; r\tau\exp\left[\frac{p_{T}u_{r}}{T}\cos(\phi-\phi_{p})-\frac{p_{T}u_{\phi}}{rT}\sin(\phi-\phi_{p})\right]\nonumber\\&&\times\left\{ m_{T}K_{1}\left(\frac{m_{T}}{T}\right)+K_{0}\left(\frac{m_{T}}{T}\right)\left[-p_{T}\cos(\phi-\phi_{p})\frac{\partial\tau}{\partial r}+\frac{p_{T}}{r}\sin(\phi-\phi_{p})\frac{\partial\tau}{\partial\phi}\right]\right\} \nonumber \\
    &\equiv& J_1 +J_2 +J_3,
\end{eqnarray}
where $K_{n}(x)$ is the modified Bessel function of the second kind
\begin{eqnarray}
K_{n}(x) & = & \int_{0}^{\infty}dy\;\exp\left[-x\cosh(y-Y)\right]\;\cosh(ny-nY).
\end{eqnarray}
Here, $J_i$ is
\begin{eqnarray}
J_{1} & = & J_{1}^{(0)}+\delta J_{1}^{(0)}a_{2}^{2}+\delta J_{1}^{(2)}a_{2}\cos(2\phi_{p})+\delta J_{1}^{(4)}a_{4}\cos(4\phi_{p})+\delta J_{1}^{(4)*}a_{2}^{2}\cos(4\phi_{p}).\nonumber \\
J_{2} & = & J_{2}^{(0)}+\delta J_{2}^{(0)}a_{2}^{2}+\delta J_{2}^{(2)}a_{2}\cos(2\phi_{p})+\delta J_{2}^{(4)}a_{4}\cos(4\phi_{p})+\delta J_{2}^{(4)*}a_{2}^{2}\cos(4\phi_{p}),\nonumber \\
J_{3} & = & \delta J_{3}^{(2)}a_{2}\cos(2\phi_{p})+\delta J_{3}^{(4)}a_{4}\cos(4\phi_{p})+\delta J_{3}^{(4)*}a_{2}^{2}\cos(4\phi_{p}).
\label{eq:J_i}
\end{eqnarray}
The explicit expression for $J_i$ is shown in Append.~\ref{sec:appendix}. 
From Eq.~(\ref{eq:Cooper_Frye_formula}), we can derive $v_2$ and $v_4$,
\begin{eqnarray}
v_{2} & = & \chi_{22}\epsilon_{2},\nonumber\\
v_{4} & = & \chi_{44}\epsilon_{4}+\chi_{42}\epsilon_{2}^{2}.\label{eq:v2_v4_01}
\end{eqnarray}
where $\chi_{ij}$ can be interpreted as the response coefficient relating $\epsilon_j$ to $v_i$,
\begin{eqnarray}
\chi_{22} & = & -\frac{1}{2}\frac{\delta J_{1}^{(2)}+\delta J_{2}^{(2)}+\delta J_{3}^{(2)}}{J_{1}^{(0)}+J_{2}^{(0)}},\nonumber\\
\chi_{44} & = & -\frac{35}{72}\frac{\delta J_{1}^{(4)}+\delta J_{2}^{(4)}+\delta J_{3}^{(4)}}{J_{1}^{(0)}+J_{2}^{(0)}},\nonumber\\
\chi_{42} & = &-\frac{\left(\delta J_{1}^{(4)}+\delta J_{2}^{(4)}+\delta J_{3}^{(4)}\right)-2\left(\delta J_{1}^{(4)*}+\delta J_{2}^{(4)*}+\delta J_{3}^{(4)*}\right)}{4(J_{1}^{(0)}+J_{2}^{(0)})}.
\label{eq:chi_ij_experssion}
\end{eqnarray}
Eqs.~(\ref{eq:v2_v4_01}) clearly show the nonlinear contribution from $\epsilon_2^2$ to $v_4$. We can also rewrite the result as
\begin{equation}
v_4=\chi_{44}\epsilon_{4}+\chi v_2^{2},\;\;\chi\equiv \chi_{42}/\chi_{22}^{2}. \label{eq:v4_v22}
\end{equation}
The nonlinear coefficient $\chi$ characterizes the contribution of $v_2^2$ to $v_4$. 
Here, we emphasize that the above definition of the nonlinear response coefficient $\chi$ differs from the one commonly used in experiments. For the discussion connected to experimental measures, we refer to Sec.~\ref{sec:different_planes}.

\begin{figure}[t]
\includegraphics[width=1.0\linewidth]{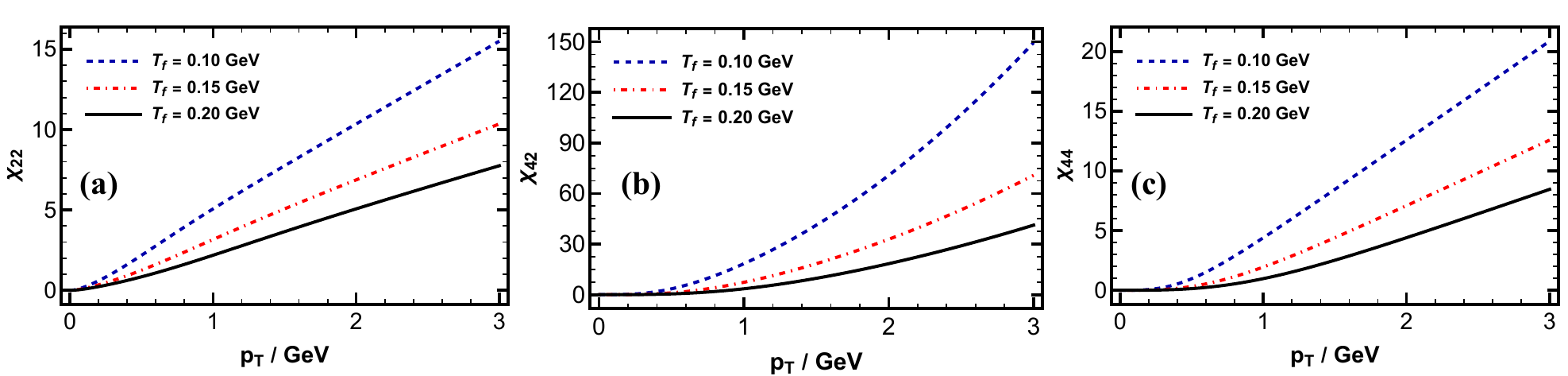}
\caption{The coefficients (a) $\chi_{22}$, (b) $\chi_{42}$ and (c) $\chi_{44}$ defined in Eq.~\eqref{eq:v2_v4_01}. We have chosen $m_0=0.1$ GeV, $L=17\;\textrm{fm}/c$ , $B=3.0$ GeV and $Y=0$. The blue dashed, red dot-dashed and black solid lines stand for the results with $T_f = 0.10, 0.15, 0.20$ GeV, respectively.}
\label{fig:chi_ij}
\end{figure}

In Fig.~\ref{fig:chi_ij}, we show $\chi_{ij}$ as a function of $p_T$ for $T_f=0.10$, $0.15$, and $0.20$ GeV at mid-rapidity ($Y=0$). As an illustrative example, we consider light hadrons and take $m_0=0.10$ GeV. 
We have also checked numerically that the $m_0$ dependence of $\chi_{ij}$ is very weak in our model.
The parameters $L$ and $B$ are chosen following Ref.~\citep{Hatta:2015era}, where they were fitted to the $v_n$ data measured by the CMS Collaboration in $\sqrt{s_{NN}}=2.76$ TeV Pb+Pb collisions with $0-0.2\%$ centrality \citep{CMS-PAS-HIN-12-011}.
We find that the magnitude of $\chi_{ij}$ is relatively large. This is because $\chi_{ij}$ scales as $1/B^a$ with $a>0$; see, for example, Eqs.~\eqref{eq:v2_v4_small_pT} in the next subsection. Consequently, smaller values of $B$ lead to larger values of $|\chi_{ij}|$.

In Fig.~\ref{fig:chi_ij}, we observe that all $\chi_{ij}$ increase with increasing $p_T$. At fixed $p_T$, $\chi_{ij}$ decrease as the freeze-out temperature $T_f$ increases. Since the expressions for $\chi_{ij}$ are algebraically cumbersome and obscure their physical interpretation, we analyze the small- and large-$p_T$ limits of these coefficients in the following subsection.

\subsection{$\chi_{ij}$ in small and large $p_{T}$ limits }\label{sec:small_pT_limits}

In small $p_T$ limit, Eqs.~(\ref{eq:J_i}) reduce to Eq.~(\ref{eq:J_small_pT}) and Eqs.~(\ref{eq:v2_v4_01}) becomes, 
\begin{eqnarray}
    \chi_{22} & = & \frac{2^{10}}{21B^{6}T_{f}^{2}}\left[1+\frac{4T_{f}K_{0}\left(\frac{m_{0}}{T}\right)}{m_{0}K_{1}\left(\frac{m_{0}}{T}\right)}\right]p_{T}^{2}+\mathcal{O}(p_{T}^{4}),\nonumber\\
\chi_{44} & = & \frac{2^{16}}{663B^{12}T_{f}^{4}}\left[7+\frac{56T_{f}K_{0}\left(\frac{m_{0}}{T}\right)}{m_{0}K_{1}\left(\frac{m_{0}}{T}\right)}\right]p_{T}^{4}+\mathcal{O}(p_{T}^{6}),\nonumber\\
\chi_{42} & = & \frac{2^{18}}{7735B^{12}T_{f}^{4}}\left[67+\frac{549T_{f}K_{0}\left(\frac{m_{0}}{T}\right)}{m_{0}K_{1}\left(\frac{m_{0}}{T}\right)}\right]p_{T}^{4}+\mathcal{O}(p_{T}^{6}).
\label{eq:v2_v4_small_pT}
\end{eqnarray}
%\begin{eqnarray}
%\chi_{22}&=&\frac{2^{10}p_{T}^{2}\left[K_{1}\left(\frac{m_{T}}{T_{f}}\right)m_{T}+4K_{0}\left(\frac{m_{T}}{T_{f}}\right)T_{f}\right]}{21B^{6}K_{1}\left(\frac{m_{T}}{T_{f}}\right)m_{T}T_{f}^{2}},\nonumber\\
%\chi_{44}&=&\frac{2^{16}p_{T}^{4}\left[7K_{1}\left(\frac{m_{T}}{T_{f}}\right)m_{T}+56K_{0}\left(\frac{m_{T}}{T}\right)T_{f}\right]}{663B^{12}K_{1}\left(\frac{m_{T}}{T_{f}}\right)m_{T}T_{f}^{4}}, \nonumber \\
%\chi_{42}&=&\frac{2^{18}p_{T}^{4}\left[67K_{1}\left(\frac{m_{T}}{T_{f}}\right)m_{T}+549K_{0}\left(\frac{m_{T}}{T_{f}}\right)T_{f}\right]}{7735B^{12}K_{1}\left(\frac{m_{T}}{T_{f}}\right)m_{T}T_{f}^{4}}.\label{eq:v2_v4_small_pT}
%\end{eqnarray}
%Note that $m_T$ also depends on $p_T$. The modified Bessel function of the second kind $K_n(x)$ does not reduce to a simple polynomial in $x$ in the $x\rightarrow 0$ limit. We therefore keep $K_n(x)$ in Eqs.~\eqref{eq:v2_v4_small_pT} rather than writing an explicit series expansion in $p_T$.
We find that $\chi_{22}\propto p_T^2$ and $\chi_{42}, \chi_{44}\propto p_T^4$ in the small $p_T$ limit. 
We have also checked numerically that Eqs.~(\ref{eq:v2_v4_small_pT}) agree with the full analytic expressions in Eqs.~\eqref{eq:chi_ij_experssion} in the small $p_T$ limit.
It is straightforward to derive $\chi$ in the same limit, 
\begin{equation}
\chi=\frac{441}{30940}[\chi_{(0)}+\chi_{(2)}p_{T}^{2}]+\mathcal{O}(p_{T}^{4}),
\label{eq:chi_small_pT}
\end{equation}
where 
\begin{eqnarray}
\chi_{(0)} & = & \frac{m_{0}K_{1}\left(\frac{m_{0}}{T_f}\right)\left[549T_{f}K_{0}\left(\frac{m_{0}}{T_f}\right)+67m_{0}K_{1}\left(\frac{m_{0}}{T_f}\right)\right]}{\left[4T_{f}K_{0}\left(\frac{m_{0}}{T_f}\right)+m_{0}K_{1}\left(\frac{m_{0}}{T_f}\right)\right]^{2}},\nonumber \\
\chi_{(2)} & = & -\frac{1}{2}\frac{\left[2196T_{f}K_{0}\left(\frac{m_{0}}{T_f}\right)-13m_{0}K_{1}\left(\frac{m_{0}}{T_f}\right)\right]\left\{ \left[K_{0}\left(\frac{m_{0}}{T_f}\right)\right]^{2}-\left[K_{1}\left(\frac{m_{0}}{T_f}\right)\right]^{2}\right\} }{\left[4T_{f}K_{0}\left(\frac{m_{0}}{T_f}\right)+m_{0}K_{1}\left(\frac{m_{0}}{T_f}\right)\right]^{3}}.
\end{eqnarray}
Interestingly, in the small $p_T$ limit, $\chi$ is independent of $B$.  
%\begin{eqnarray}
%\chi&=&\frac{441}{30940}\frac{\left[67K_{1}\left(\frac{m_{T}}{T_{f}}\right)m_{T}+549K_{0}\left(\frac{m_{T}}{T_{f}}\right)T\right]K_{1}\left(\frac{m_{T}}{T_{f}}\right)m_{T}}{\left[K_{1}\left(\frac{m_{T}}{T_{f}}\right)m_{T}+4K_{0}\left(\frac{m_{T}}{T_{f}}\right)T\right]^{2}}.
%\end{eqnarray}
%\begin{eqnarray}
%\chi&=&\frac{441(67K_{1}m_{T}+549K_{0}T)K_{1}m_{T}}{30940(K_{1}m_{T}+4K_{0}T)^{2}}.
%\end{eqnarray}
%We could see it is consistent with Fig.~\ref{fig:chi}(a) that $\chi$ increases in small $p_{T}$ limit. 

\begin{figure}[t]
\includegraphics[width=1.0\linewidth]{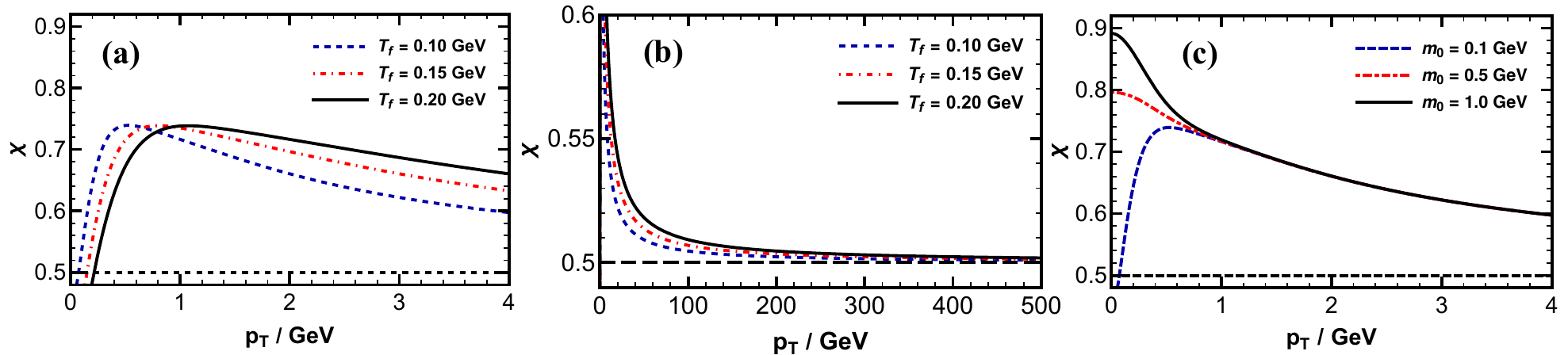}
\caption{The ratio $\chi(p_{T})$, defined in Eq.(\ref{eq:v4_v22}), as a function of $p_T$. Panels (a) and (b) show $\chi$ in the small $p_T$ and large $p_T$ regions, respectively, for different freeze-out temperatures $T_f$ at fixed $m_0=0.1$ GeV. Panel (c) shows $\chi(p_{T})$ for different $m_0$ at fixed $T_f=0.10$ GeV. The other parameters are chosen to be the same as those in Fig.~\ref{fig:chi_ij}.}
\label{fig:chi}
\end{figure}

Similarly, we can analyze $\chi_{ij}$ in the large-$p_T$ limit. Using the saddle-point approximation, we find that, for large $p_T$,
\begin{eqnarray}
v_2 &\approx& \frac{p_{T}}{T}\frac{500\sqrt{5}}{81\;B^{3}}\epsilon_{2}, \nonumber \\
\frac{v_4}{v_2^2} &=& \frac{1}{2} +\mathcal{O}(p_T^{-1}). \label{eq:v_4_large_pT}
\end{eqnarray}
More details of the derivation are provided in Appendix~\ref{sec:appendix_limits}. 
In early works~\citep{Borghini:2005kd, Teaney:2012ke, Lang:2013oba, Gombeaud:2009ye} , it was pointed out that $v_{4}/v_{2}^{2}\rightarrow1/2$ is universal for ideal fluids. Our result provides an analytical confirmation of this behavior within the Gubser model.

In Fig.~\ref{fig:chi}, we show $\chi$ as a function of $p_T$. The magnitude of $\chi$ is consistent with the data reported in Ref.~\citep{CMS:2013wjq}.
In Fig.~\ref{fig:chi}(a), we find that $\chi$ exhibits only a weak dependence on the freeze-out temperature $T_f$. We also observe a maximum at intermediate transverse momentum, $p_T\sim 0.5 - 1.0$ GeV. In Fig.~\ref{fig:chi}(b), $\chi$ approaches $1/2$ in the large $p_T$ limit, consistent with Eq.~\eqref{eq:v_4_large_pT}. 
We further find a nontrivial dependence of $\chi$ on $m_0$ in Fig.~\ref{fig:chi}(c), in contrast to $\chi_{ij}$, which depends only weakly on $m_0$. For small $m_0$, $\chi$ increases with $p_T$ in the small $p_T$ region, whereas for sufficiently large $m_0$ it decreases monotonically with $p_T$. Correspondingly, the intermediate-$p_T$ maximum in Fig.~\ref{fig:chi}(a) disappears when $m_0$ is sufficiently large. 
We have also checked numerically that Eq.\eqref{eq:chi_small_pT} is consistent with the results in Fig.\ref{fig:chi}(c) for $p_T<0.2$ GeV.
In the large-$p_T$ region, $\chi$ computed with different $m_0$ values converges to the same asymptotic behavior.

\begin{figure}[t]
    \centering
    \includegraphics[width=0.4\linewidth]{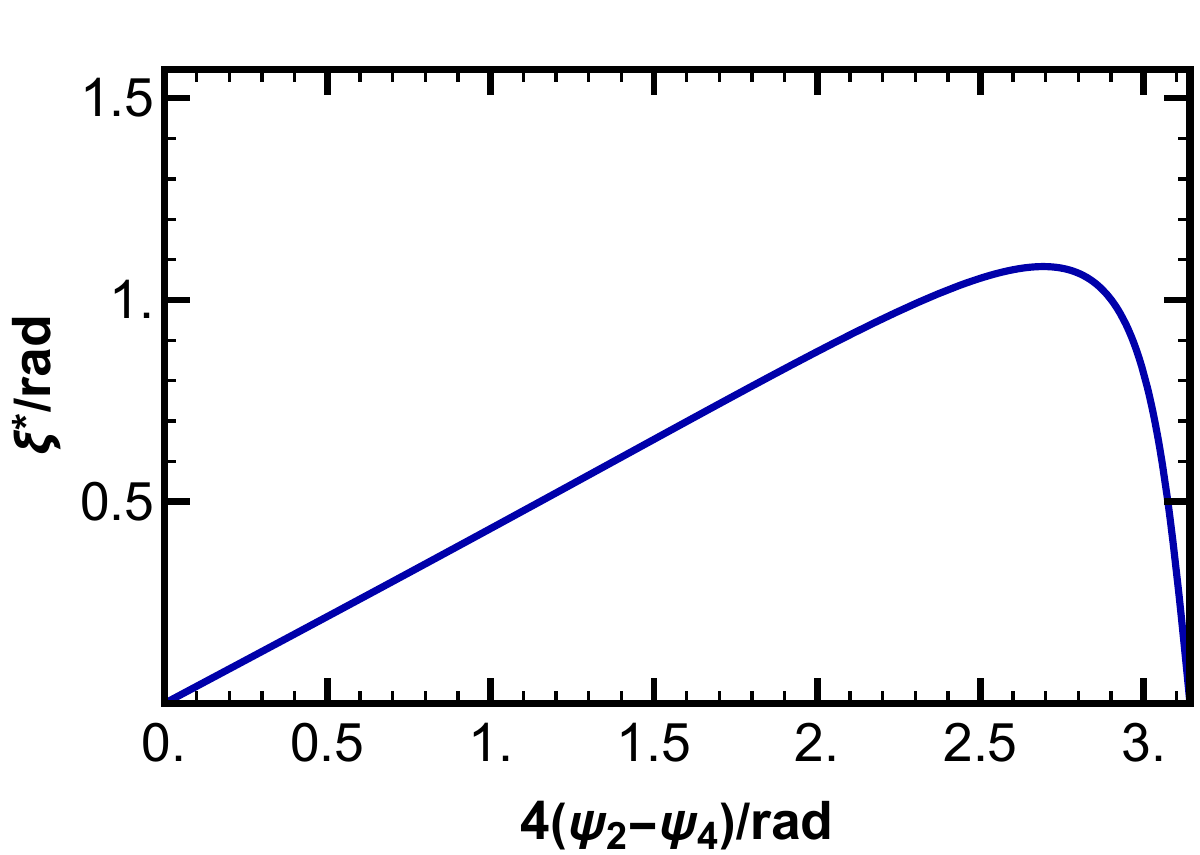}
    \caption{Angle $\xi^*$ as a function of $4(\psi_2-\psi_4)$. The parameters are chosen to be the same as those in Fig.~\ref{fig:chi_ij}, with $\epsilon_2=0.04$ and $\epsilon_4=0.01$.}
    \label{fig:xi}
\end{figure}

%\section{Effects of the mismatch between the participant and reaction planes \haojie{Connection to the experimental measurements}} \label{sec:different_planes}
\section{Connection to the experimental measurements}
\label{sec:different_planes}

In the previous section, we assumed that the perturbation $S(\theta,\phi)$ follows Eq.~\eqref{eq:S_theta_phi}. In realistic event-by-event hydrodynamics, the participant-plane angle $\psi_n$ associated with the initial-state eccentricity $\epsilon_n$ can differ from the reaction-plane angle $\Psi_n$. To better quantify anisotropic flow, we therefore need to account for the mismatch between $\psi_n$ and $\Psi_n$ and its impact on the relation between $\epsilon_n$ and the measured flow harmonics $v_n$.

We now rewrite the perturbation $S(\theta,\phi)$ to include the finite participant-plane angles $\psi_n$,
\begin{eqnarray}
S(\theta,\phi) & = & a_{2}\left(\frac{2rL}{L^{2}+r^{2}}\right)^{2}\cos[2(\phi-\psi_{2})]+a_{4}\left(\frac{2rL}{L^{2}+r^{2}}\right)^{4}\cos[4(\phi-\psi_{4})].
\end{eqnarray}
Note that this modification of $S$ does not affect the general solutions for the velocity perturbations in Eqs.~(\ref{eq:General_solution_nu}) or the other perturbation parameters in Eqs.~(\ref{eq:perturbation_parameters}). 
Then, following the same processes in Sec.~\ref{sec:EccetricityV2V4}, we find the Eqs.~\eqref{eq:Relation_eccentricity} becomes up to $\mathcal{O}(a_2a_4)$, 
\begin{eqnarray}
\epsilon_{2} & = & -a_{2},\nonumber \\
\epsilon_{4} & = & -\frac{36}{35}a_{4}-\frac{18}{35}\cos[4(\psi_{2}-\psi_{4})]a_{2}^{2}.
\end{eqnarray}
Or, we can write the above equations as,
\begin{eqnarray}
a_{2} & = & -\epsilon_{2},\nonumber \\
a_{4} & = & -\frac{35}{36}\epsilon_{4}-\frac{1}{2}\cos[4(\psi_{2}-\psi_{4})]\epsilon_{2}^{2}.\label{eq:New_epsilon}
\end{eqnarray}
Compared with the relationship in Eq. (\ref{eq:Relation_eccentricity}),
an additional factor $\cos [4(\psi_4 - \psi_2)]$ appears in $\epsilon_4$.

Then the Cooper-Frye formula incorporates event planes $\Psi_{n}$,
\begin{eqnarray}
(2\pi)^{3}\frac{dN}{dYp_{T}dp_{T}d\phi_{p}} & = & -\int_{\Sigma}p^{\mu}d\sigma_{\mu}f(p^{\mu}u_{\mu}/T)\propto1+2v_{n}(p_{T})\cos[n(\phi_{p}-\Psi_{n})].\label{eq:CF_new}
\end{eqnarray}
Using the same parameterization of the freeze-out temperature $T_f$ as in Eq.~\eqref{eq:T_f}, we obtain the modified freeze-out proper time $\tau_f$ and the transverse flow components $u_r$ and $u_\phi$,
\begin{eqnarray}
\tau & = & \tau_{0}+\delta\tau_{0}a_{2}^{2}+\delta\tau_{2}a_{2}\cos[2(\phi-\psi_{2})]+\delta\tau_{4(1)}a_{4}\cos[4(\phi-\psi_{4})]+\delta\tau_{4(2)}a_{2}^{2}\cos[4(\phi-\psi_{2})],\nonumber \\
u_{\phi} & = & \delta u_{\phi2}a_{2}\sin[2(\phi-\psi_{2})]+\delta u_{\phi4(1)}a_{4}\sin[4(\phi-\psi_{4})]+\delta u_{\phi4(2)}a_{2}^{2}\sin[4(\phi-\psi_{2})],\nonumber \\
u_{r} & = & u_{r0}+\delta u_{r0}a_{2}^{2}+\delta u_{r2}a_{2}\cos[2(\phi-\psi_{2})]+\delta u_{r4(1)}a_{4}\cos[4(\phi-\psi_{4})]+\delta u_{r4(2)}a_{2}^{2}\cos[4(\phi-\psi_{2})].
\label{eq:tau_f_new}
\end{eqnarray}
The coefficients $\tau_{0},\delta \tau_{0},\delta\tau_{2},\delta\tau_{4(1)},\delta\tau_{4(2)}, \delta u_{\phi2}, \delta u_{\phi4(1)},\delta u_{\phi4(2)}, u_{r0}, \delta u_{r0}, \delta u_{r2}, \delta u_{r4(1)}$ and $\delta u_{r4(2)}$ are the same as those in Eqs.~(\ref{eq:Freezeout_time}, \ref{eq:velocity_freezeout}).
Finally, we obtain that, 
\begin{eqnarray}
v_{2}\cos2(\phi_{p}-\Psi_{2}) & = & \chi_{22}\epsilon_{2}\cos2(\phi_{p}-\psi_{2}),\nonumber \\
v_{4}\cos4(\phi_{p}-\Psi_{4}) & \approx & \{\chi_{44}\epsilon_{4}+\chi_{42}\text{cos}[4(\psi_{4}-\psi_{2})]\epsilon_{2}^{2}\}\cos[4(\phi_{p}-\psi_{4})+\xi^{*}],\label{eq:v2_v4_new}
\end{eqnarray}
where $\chi_{22},\chi_{44}$ and $\chi_{42}$ are the same as those
in Eqs.~(\ref{eq:chi_ij_experssion}), and 
\begin{equation}
\text{tan}\xi^{*}=\frac{36G_{2}\sin[4(\psi_{4}-\psi_{2})]\epsilon_{2}^{2}}{-35G_{1}\epsilon_{4}+36(G_{2}-\frac{1}{2}G_{1})\epsilon_{2}^{2}\cos[4(\psi_{4}-\psi_{2})]},
\label{eq:xi}
\end{equation}
with $G_{1}=\delta J_{1}^{(4)}+\delta J_{2}^{(4)}+\delta J_{3}^{(4)}$,
and $G_{2}=\delta J_{1}^{(4)*}+\delta J_{2}^{(4)*}+\delta J_{3}^{(4)*}$.
By comparing both sides of Eq.~(\ref{eq:v2_v4_new}), we find 
\begin{eqnarray}
\Psi_{2} & = & \psi_{2},\nonumber \\
\Psi_{4} & = & \psi_{4}-\frac{1}{4}\xi^{*}+2\pi k,
\label{eq:Psi_psi}
\end{eqnarray}
where $k\in\mathbb{Z}$ ensures that $\Psi_{4}\in[0,2\pi)$.
Equation~\eqref{eq:Psi_psi} shows that once the initial angles $\psi_2$ and $\psi_4$ are specified, $\Psi_4$ is not arbitrary.
These relations also imply that, in general, $\Psi_{4}\neq\psi_{4}$. Only when $\psi_{4}=\psi_{2}+\pi k$ with $k\in\mathbb{Z}$ does $\xi^{*}$ vanish, yielding $\Psi_{4}=\psi_{4}$. 
We also plot the angle $\xi^{*}$ as a function of $4(\psi_2-\psi_4)$ in Fig.~\ref{fig:xi}. Owing to the $\sin$ function in the numerator on the right-hand side of Eq.~\eqref{eq:xi}, $\xi^{*}$ is non-monotonic and vanishes when $4(\psi_2-\psi_4)=0,\pi$.

%{\color{blue} The following part will be moved to the end of this section.}
%\haojie{Analogously, the $v_{4}$ can be written as
%\begin{equation}
%v_{4}=\chi_{44}\epsilon_{4}+(\chi_{44}/\chi_{22}^{2})\text{cos}[4(\psi_{4}-\psi_{2})]v_{2}^{2}.
%\end{equation}
%In previous studies, $v_{4}$ was written as 
%\begin{equation}
%v_{4} = v_{4,\rm NL} + \chi v_{2}^{2}
%\end{equation}
%with
%\begin{equation}
%\chi = \frac{\langle v_{4}v_{2}^{2}\cos[4(\Psi_{4}-\Psi_2)]\rangle}{\langle v_{2}^{4} \rangle}
%\end{equation}
%The $\langle ...\rangle$ indicate the average over ensemble of events. We do not include the event fluctuation in the analytical derivation, therefore, 
%\begin{equation}
%\chi = \frac{v_{4}\cos[4(\Psi_{4}-\Psi_2)]}{ v_{2}^{2}}
%\end{equation}
%}

We can still introduce the $\chi=\chi_{42}/\chi_{22}^2$, such that
\begin{equation}
v_{4}=\chi_{44}\epsilon_{4}+\chi\text{cos}[4(\psi_{4}-\psi_{2})]v_{2}^{2}.
\end{equation}
The corresponding effective nonlinear coefficient is then $\chi\cos[4(\psi_{4}-\psi_{2})]$.
To facilitate comparison with the standard experimental treatment (e.g. see Refs.~\citep{Gardim:2014tya,Yan:2015jma,Luzum:2011mm,ALICE:2011ab,PHENIX:2011yyh}), we introduce two vectors in the complex plane,
\begin{eqnarray}
V_{2} & = & v_{2}\{\Psi_{2}\}e^{i2\Psi_{2}}=v_{2}\{\Psi_{2}\}e^{i2\psi_{2}},
\end{eqnarray}
where we have used Eq. \eqref{eq:Psi_psi}, and
\begin{equation}
V_{4}=v_{4}\{\Psi_{4}\}e^{i4\Psi_{4}}.
\end{equation}
Here, $v_{n}\{\Psi_{m}\}$ denotes the $n$-th harmonic flow coefficient measured with respect to the reaction plane direction $\Psi_{m}$.

For $V_{4}$, it is common to project it onto the $\psi_{n}$ plane. Following Eq.~\eqref{eq:Psi_psi}, $V_{4}$ can be decomposed into two parts,
\begin{equation}
V_{4}=v_{4L}e^{i4\psi_{4}}+\chi v_{2}^{2}e^{4i\Psi_{2}}\equiv v_{4L}e^{4i\psi_{4}}+v_{4}\{\Psi_{2}\}e^{4i\Psi_{2}}.
\end{equation}
After some algebra, we obtain
\begin{eqnarray}
v_{4}\{\Psi_{4}\} & = & \sqrt{v_{4L}^{2}+2v_{4L}v_{4}\{\Psi_{2}\}\cos4(\psi_{2}-\psi_{4})+v_{4}^{2}\{\Psi_{2}\}}.\label{eq:v4_Psi_01}
\end{eqnarray}
We can further expand the above equation in the limits of small $v_4\{\Psi_2\}$ or small $v_{4L}$.
In our power-counting scheme, $v_4\{\Psi_2\}/v_{4L}\ll1$, Eq.~\eqref{eq:v4_Psi_01}
\begin{eqnarray}
    v_{4}\{\Psi_{4}\}&\approx&v_{4L}+v_{4}\{\Psi_{2}\}\cos[4(\psi_{2}-\psi_{4})]=v_{4L}+\chi\cos[4(\psi_{2}-\psi_{4})]v_{2}\{\Psi_{2}\}^{2}.
    \label{eq:v4_Psi_4}
\end{eqnarray}

We now focus on Eq.~(\ref{eq:v4_Psi_4}), which is a main highlight of this work.
As commonly assumed in both experimental analyses and theoretical calculations (see, e.g., Refs.~\citep{Yan:2015jma,Borghini:2005kd,Bravina:2013ora,Bravina:2013xla}), the factor $\cos[4(\psi_{2}-\psi_{4})]$ has been treated merely as statistical noise and therefore not examined in detail.  %In that approximation, Eq.~(\ref{eq:v4_Psi_4}) is written as $v_{4}\{\Psi_{4}\}=v_{4L}+\chi v_{2}\{\Psi_{2}\}^{2}$. 
Here, our result makes explicit the additional dependence on the initial participant-plane angles. 

\begin{figure}[t]
    \centering
\includegraphics[width=0.7\linewidth]{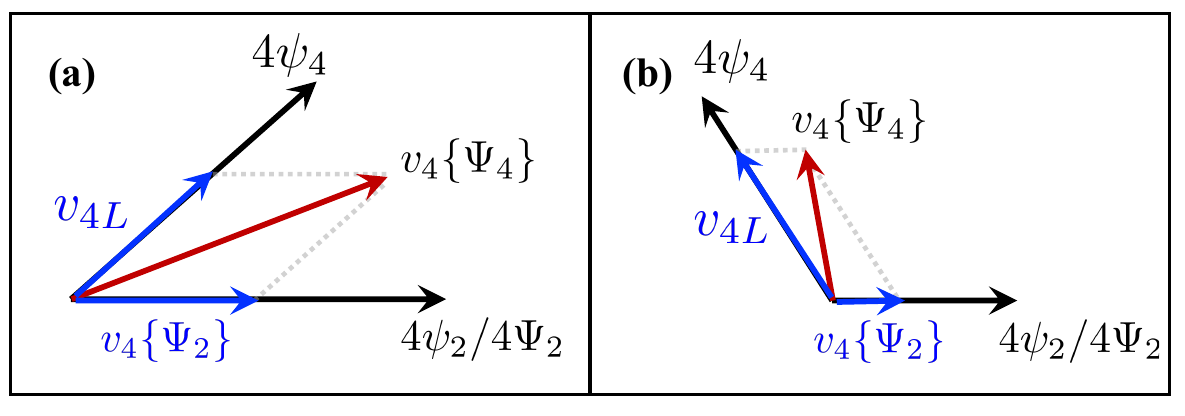}
    \caption{Illustration of Eq.~(\ref{eq:v4_Psi_01}, \ref{eq:v4_Psi_4}) for (a) $\cos[4(\psi_{2}-\psi_{4})]>0$ and (b) $\cos[4(\psi_{2}-\psi_{4})]<0$. }
    \label{fig:v4_Psi4}
\end{figure}

Let us discuss Eq.\eqref{eq:v4_Psi_4} in four different cases. 
\begin{itemize}
    \item Case I: When $4(\psi_{2}-\psi_{4})=0$, i.e. $\cos[4(\psi_{2}-\psi_{4})]= 1$, the angle $\xi^*$ in Eq.\eqref{eq:xi} also vanishes. In this case, $\Psi_4 =\psi_4=\psi_2$, and Eq.~(\ref{eq:v4_Psi_4}) reduces to the commonly used relation between $v_{4}\{\Psi_{4}\}$ and $v_{4}\{\Psi_{2}\}$, i.e. $v_{4}\{\Psi_{4}\}=v_{4L}+\chi v_{2}\{\Psi_{2}\}^{2}$.
    \item Case II: When $4(\psi_{2}-\psi_{4})\in(0,\pi/2)$, i.e. $\cos[4(\psi_{2}-\psi_{4})]>0$, the effective nonlinear coefficient is positive and $v_4\{\Psi_4\}>v_{4L}$. See Fig.~\ref{fig:v4_Psi4}(a) for an example.
    \item Case III: When $4(\psi_{2}-\psi_{4})=\pi/2$, i.e. $\cos[4(\psi_{2}-\psi_{4})]=0$, Eq.~\eqref{eq:v4_Psi_01} yields the simple relation $v_4\{\Psi_4\}^2 = v_{4L}^2 + v_4\{\Psi_2\}^2= v_{4L}^2 + \chi^2 v_2\{\Psi_2\}^4$. In this case, the quartic term $v_2\{\Psi_2 \}^4$ appears instead of the quadratic term $v_2\{\Psi_2 \}^2$, i.e. the usual effective nonlinear coefficient vanishes.
    \item Case IV: When $4(\psi_{2}-\psi_{4})\in(\pi/2,\pi)$, i.e. $\cos[4(\psi_{2}-\psi_{4})]<0$, the effective nonlinear coefficient becomes negative and $v_4\{\Psi_4\}<v_{4L}$. See Fig.~\ref{fig:v4_Psi4}(b) for an example.
\end{itemize}

More generally, the opposite hierarchy $v_{4L}/v_4\{\Psi_2\}\ll1$ is also possible. In that case, Eq.~\eqref{eq:v4_Psi_01} becomes 
\begin{eqnarray}
    v_{4}\{\Psi_{4}\}\approx v_{4L}\cos[4(\psi_{2}-\psi_{4})]+v_{4}\{\Psi_{2}\}=v_{4L}\cos[4(\psi_{2}-\psi_{4})]+\chi v_{2}\{\Psi_{2}\}^{2},
    \label{eq:v4_Psi_02}
\end{eqnarray}
The discussion for this case is similar to that for Eq.~\eqref{eq:v4_Psi_4}. 

Finally, we comment on the large $p_T$ limit of Eq.\eqref{eq:v_4_large_pT}. In this limit, the effective linear contribution scales as $v_{4L}\sim \mathcal{O}(p_T^{-1})$, which is parametrically smaller than the nonlinear term $v_4\{\Psi_2\}\sim \mathcal{O}(1)$. Eq.~\eqref{eq:v4_Psi_02} then implies $v_4/v_2^2 \rightarrow 1/2$ still holds at large $p_T$.

Before closing this section, we comment on how our analytical findings relate to the standard experimental extraction of the nonlinear response, and how this connects to the recent interest in probing initial-state nuclear structure. 

In experimental measurements, the initial participant-plane angles $\psi_n$ are inaccessible on an event-by-event basis. Consequently, experimentalists construct estimators using only the measurable final-state reaction planes, defining the effective nonlinear coefficient via an ensemble average~\citep{Yan:2015jma,Borghini:2005kd,Bravina:2013ora,Bravina:2013xla}:
\begin{equation}
\chi_{\textrm{N}} = \frac{\langle v_{4}v_{2}^{2}\cos[4(\Psi_{4}-\Psi_2)]\rangle}{\langle v_{2}^{4} \rangle}.
\label{eq:chi_N}
\end{equation}
By relying on this statistical average over the final planes, the standard approach implicitly assumes an orthogonal decomposition. It essentially treats the underlying geometric mismatch between the initial spatial planes as statistical noise that cleanly averages out, leaving $\chi_{\textrm{N}}$ to be interpreted almost exclusively as a reflection of the QGP medium's transport properties (e.g., shear viscosity).

However, our exact analytical derivation in Eq.~\eqref{eq:v4_Psi_4} demonstrates that the geometric phase mismatch $\cos[4(\psi_2 - \psi_4)]$ is an intrinsic, non-removable mathematical feature of the hydrodynamic mapping. This analytical realization provides crucial theoretical underpinning for recent phenomenological proposals that the nonlinear response coefficient is sensitive to the initial-state nuclear deformation, particularly the hexadecapole deformation ($\beta_4$)~\citep{Xu:2024bdh}.

In collisions of highly deformed nuclei, the intrinsic structural shapes ($\beta_2$ and $\beta_4$) strongly correlate the initial spatial eccentricities $\epsilon_2$ and $\epsilon_4$. More importantly, this structural deformation dictates the preferred relative orientations of the initial participant planes $\psi_2$ and $\psi_4$. Because of this shape-driven correlation, the projection factor $\cos[4(\psi_2 - \psi_4)]$ does not simply average to a trivial background; instead, it retains a macroscopic "memory" of the initial nuclear geometry. 

Therefore, our analytical result proves that when experimentalists measure $\chi_{\textrm{N}}$, they are not extracting a pure medium response. Rather, they are extracting a convolution of the true dynamic medium response ($\chi$) and the geometric projection factor ($\cos[4(\psi_2 - \psi_4)]$) governed by nuclear structure. By isolating this geometric phase difference within the exact Gubser flow framework, our study establishes a rigorous mathematical foundation for why higher-order nonlinear flow correlations can serve as precision tools to image the intrinsic shapes of colliding nuclei. A more detailed, event-by-event phenomenological analysis quantifying these specific deformation effects is beyond the scope of the present theoretical work, and will be presented elsewhere.

\section{Summary \label{sec:Conclusion}}

In this work, we have studied the nonlinear relation between the harmonic flow coefficients $v_2,v_4$ and the eccentricities $\epsilon_2,\epsilon_4$, as well as the effects of a mismatch between the participant and reaction planes, within the Gubser-flow framework.
We first extend the perturbative solutions of Gubser flow derived in Refs.~\citep{Hatta:2014upa,Hatta:2014jva,Hatta:2015era,Hatta:2015hca} and obtain the response coefficients $\chi_{ij}$, as shown in Eqs.~\eqref{eq:v2_v4_01}. We also define the nonlinear coefficient $\chi$ in Eq.~\eqref{eq:v4_v22} and find that $v_4/v_2^2\simeq\chi\rightarrow 1/2$ in the large $p_T$ limit, in agreement with previous studies~\citep{Borghini:2005kd, Teaney:2012ke, Lang:2013oba, Gombeaud:2009ye}. 

Next, we consider the effects of a mismatch between the participant and reaction planes. We find that the effective nonlinear response coefficient becomes $\chi \cos[4(\psi_2 -\psi_4)]$, which can vanish or become negative. We also decompose the $v_4\{\Psi_4\}$ as in Eq.~\eqref{eq:v4_Psi_4}. Again, the additional factor $\cos[4(\psi_2 -\psi_4)]$ appears. 
%This differs from the commonly used decomposition of $v_4\{\Psi_4\}$.
This additional factor can modify both the strength but even the sign of the effective nonlinear response coefficient, making it sensitive to the initial configuration of the colliding nuclei.
On the other hand, we find that $v_4/v_2^2 \rightarrow 1/2$ still holds at large $p_T$.
Our finding extends the commonly used treatment of $v_4\{\Psi_4\}$, and a detailed phenomenological study will be presented elsewhere.

Before concluding, we comment on dissipative effects. In the present work, we focus on an ideal Gubser flow for simplicity. According to Refs.~\citep{Hatta:2014upa,Hatta:2014jva,Hatta:2015era,Hatta:2015hca}, shear viscosity modifies the response coefficients $\chi_{ij}$. We therefore expect that the nonlinear response coefficient $\chi$ will also be modified once dissipative effects are included. By contrast, the additional factor $\cos[4(\psi_2 -\psi_4)]$, which arises from the mismatch between $\psi_2$ and $\psi_4$, is expected to remain. Our main conclusions should therefore persist in dissipative hydrodynamics.

\begin{acknowledgments}
This work is supported in part by the National Key Research
and Development Program of China under Contract No. 2022YFA1605500,
by the Chinese Academy of Sciences (CAS) under Grant No. YSBR-088 and by National Natural Science Foundation of China (NSFC) under Grants No.~12135011, No.~12275082.
\end{acknowledgments}

\appendix

\section{Expression for $\delta u_r$, $\delta u_\phi$ in Eq.~(\ref{eq:velocity_freezeout}) and $J_i$ in Eq.~(\ref{eq:J_i})} \label{sec:appendix}

The explicit expressions for $\delta u_{\phi2}$, $\delta u_{\phi4}$, $\delta u_{r2}$, $u_{r0}$, and $\delta u_{r4}$ in Eq.~(\ref{eq:velocity_freezeout}) are shown as follow,
\begin{eqnarray}
\delta u_{\phi2} & = & \frac{(2L)^{5}}{B^{3}(L^{2}+r^{2})^{2}}\times3\left(\frac{2rL}{L^{2}+r^{2}}\right)^{2},\nonumber \\
\delta u_{\phi4(1)} & = & \frac{(2L)^{5}}{B^{3}(L^{2}+r^{2})^{2}}\times6\left(\frac{2rL}{L^{2}+r^{2}}\right)^{4},\nonumber \\
\delta u_{\phi4(2)} & = & \frac{(2L)^{5}}{B^{3}(L^{2}+r^{2})^{2}}\times\frac{9}{2}\left(\frac{2rL}{L^{2}+r^{2}}\right)^{4},\nonumber \\
u_{r0} & = & \frac{(2L)^{5}}{B^{3}(L^{2}+r^{2})^{2}}\times\frac{2r}{L^{2}+r^{2}},\nonumber \\
\delta u_{r0} & = & \frac{(2L)^{5}}{B^{3}(L^{2}+r^{2})^{2}}\times3\left(\frac{2rL}{L^{2}+r^{2}}\right)^{4}\frac{1}{L^{2}+r^{2}}\left(\frac{5r^{2}-3L^{2}}{2r}\right),\nonumber \\
\delta u_{r2} & = & \frac{(2L)^{5}}{B^{3}(L^{2}+r^{2})^{2}}\times6\left(\frac{3r^{2}-L^{2}}{2r}\right)\frac{1}{L^{2}+r^{2}}\left(\frac{2rL}{L^{2}+r^{2}}\right)^{2},\nonumber \\
\delta u_{r4(1)} & = &\frac{(2L)^{5}}{B^{3}(L^{2}+r^{2})^{2}}\times6\left(\frac{2r^{2}-L^{2}}{r}\right)\left(\frac{2rL}{L^{2}+r^{2}}\right)^{4}\frac{1}{L^{2}+r^{2}},\nonumber \\
\delta u_{r4(2)} & = & \frac{(2L)^{5}}{B^{3}(L^{2}+r^{2})^{2}}\times3\left(\frac{5r^{2}-3L^{2}}{2r}\right)\left(\frac{2rL}{L^{2}+r^{2}}\right)^{4}\frac{1}{L^{2}+r^{2}}. \label{eq:delta_u_01}
\end{eqnarray}

The expression for $\delta J^{(j)}_i$ in Eq.~(\ref{eq:J_i}) are shown as follow.
For $J_1$, we have,
\begin{eqnarray}
    J_{1}^{(0)}&=&4\pi m_{T}K_{1}\left(\frac{m_{T}}{T}\right)\int dr\  r\tau_{0}I_{0},\nonumber \\
    \delta J_{1}^{(0)}&=&4\pi m_{T}K_{1}\left(\frac{m_{T}}{T}\right)\int dr\  r\left[\delta\tau_{0}I_{0}+\tau_{0}\frac{p_{T}}{2T}\left(\frac{\delta\tau_{2}}{\tau_{0}}\delta u_{r2}+2\delta u_{r0}\right)\right]\nonumber\\
    &&+4\pi m_{T}K_{1}\left(\frac{m_{T}}{T}\right)\int_{\Sigma}dr\  r\tau_{0}\left(\frac{p_{T}}{2T}\right)^{2}\left[\delta u_{r2}^{2}\left(I_{0}-\frac{I_{1}}{z}\right)+(\frac{\delta u_{\phi2}}{r})^{2}\frac{I_{1}}{z}\right],\nonumber\\
    \delta J_{1}^{(2)}&=&4\pi m_{T}K_{1}\left(\frac{m_{T}}{T}\right)\int dr\ r\tau_{0}\left\{ \frac{\delta\tau_{2}}{\tau_{0}}I_{2}+\frac{p_{T}}{2T}\left[\left(\delta u_{r2}-\frac{\delta u_{\phi2}}{r}\right)I_{1}+\left(\delta u_{r2}+\frac{\delta u_{\phi2}}{r}\right)I_{3}\right]\right\} ,\nonumber
    \\
    \delta 
    J_{1}^{(4)}&=&4\pi m_{T}K_{1}\left(\frac{m_{T}}{T}\right)\int dr\  r\tau_{0}\left\{ \frac{\delta\tau_{4(1)}}{\tau_{0}}I_{4}+\frac{p_{T}}{2T}\left[\left(\delta u_{r4(1)}-\frac{\delta u_{\phi4(1)}}{r}\right)I_{3}+\left(\delta u_{r4(1)}+\frac{\delta u_{\phi4(1)}}{r}\right)I_{5}\right]\right\},
      \end{eqnarray}
    and,
    \begin{eqnarray}
    \delta J_{1}^{(4)*}&=&4\pi m_{T}K_{1}\left(\frac{m_{T}}{T}\right)\int dr\  r\tau_{0}\left\{ \frac{\delta\tau_{4(2)}}{\tau_{0}}I_{4}+\frac{p_{T}}{2T}\left[\left(\delta u_{r4(2)}-\frac{\delta u_{\phi4(2)}}{r}\right)I_{3}+\left(\delta u_{r4(2)}+\frac{\delta u_{\phi4(2)}}{r}\right)I_{5}\right]\right\}\nonumber\\&&+4\pi m_{T}K_{1}\left(\frac{m_{T}}{T}\right)\int dr\  r\tau_{0}\frac{p_{T}}{2T}\left(\frac{\delta\tau_{2}}{\tau_{0}}\delta u_{r2}I_{4}'-\frac{\delta\tau_{2}}{\tau_{0}}\frac{\delta u_{\phi2}}{r}\frac{4}{z}I_{4}\right)\nonumber\\&&+4\pi m_{T}K_{1}\left(\frac{m_{T}}{T}\right)\int_{\Sigma}dr\ \tau_{0}\left(\frac{p_{T}}{2T}\right)^{2}\frac{1}{2}\left(\delta u_{r2}^{2}+(\frac{\delta u_{\phi2}}{r})^{2}\right)\left(I_{3}'+I_{5}'\right)\nonumber\\&&-4\pi m_{T}K_{1}\left(\frac{m_{T}}{T}\right)\int_{\Sigma}dr\ r\tau_{0}\left(\frac{p_{T}}{2T}\right)^{2}\left[(\frac{\delta u_{\phi2}}{r})^{2}I_{4}+\delta u_{r2}\frac{\delta u_{\phi2}}{r}\left(I_{3}'-I_{5}'\right)\right].
    \label{eq:delta_J_1}
\end{eqnarray}
For convenience, we have omitted all variables in the Bessel function $I_n(z)$ with $z\equiv p_{T}u_{r0}/T$. And the definition of Bessel function is,
\begin{eqnarray}
    2\pi I_{n}(z)\cos(n\phi_{p})=\int d\phi\exp\left[z\cos(\phi-\phi_{p})\right]\cos(n\phi),
\end{eqnarray}
For $J_2$, we have,
\begin{eqnarray}
    J_{2}^{(0)}&=&-4\pi p_{T}K_{0}\left(\frac{m_{T}}{T}\right)\int dr\  r\tau_{0}\frac{\partial\tau_{0}}{\partial r}I_{1},\nonumber\\
    \delta J_{2}^{(0)}&=&-4\pi p_{T}K_{0}\left(\frac{m_{T}}{T}\right)\int dr\  r\left(\frac{1}{2}\tau_{0}\frac{\delta\tau_{2}}{\tau_{0}}\frac{\partial\delta\tau_{2}}{\partial r}+\tau_{0}\frac{\partial\delta\tau_{0}}{\partial r}+\delta\tau_{0}\frac{\partial\tau_{0}}{\partial r}\right)I_{1}\nonumber\\&&-4\pi p_{T}K_{0}\left(\frac{m_{T}}{T}\right)\int dr\ r\tau_{0}\frac{p_{T}}{2T}\left[\left(\delta u_{r2}\frac{\partial\delta\tau_{2}}{\partial r}+\delta u_{r2}\frac{\delta\tau_{2}}{\tau_{0}}\frac{\partial\tau_{0}}{\partial r}+2\frac{\partial\tau_{0}}{\partial r}\delta u_{r0}\right)\left(I_{0}-\frac{I_{1}}{z}\right)-\frac{\delta\tau_{2}}{r}\frac{\delta u_{\phi2}}{r}\frac{2I_{1}}{z}\right]\nonumber\\&&-4\pi p_{T}K_{0}\left(\frac{m_{T}}{T}\right)\int dr\  r\tau_{0}\left(\frac{p_{T}}{2T}\right)^{2}\frac{\partial\tau_{0}}{\partial r}\left[\delta u_{r2}^{2}\left(I_{1}-\frac{I_{2}}{z}\right)+(\frac{\delta u_{\phi2}}{r})^{2}\frac{I_{2}}{z}\right],\nonumber\\
    \delta J_{2}^{(2)}&=&-4\pi p_{T}K_{0}\left(\frac{m_{T}}{T}\right)\int dr\ r\tau_{0}\left(\frac{\partial\delta\tau_{2}}{\partial r}+\frac{\partial\tau_{0}}{\partial r}\frac{\delta\tau_{2}}{\tau_{0}}\right)I_{2}^{\prime}\nonumber\\&&-4\pi p_{T}K_{0}\left(\frac{m_{T}}{T}\right)\int dr\ r\tau_{0}\frac{p_{T}}{2T}\frac{\partial\tau_{0}}{\partial r}\left[\delta u_{r2}\left(I_{1}^{\prime}+I_{3}^{\prime}\right)-\frac{\delta u_{\phi2}}{r}\left(I_{1}^{\prime}-I_{3}^{\prime}\right)\right]\nonumber\\
    \delta J_{2}^{(4)}&=&4\pi p_{T}K_{0}\left(\frac{m_{T}}{T}\right)\int dr\  r\tau_{0}\frac{p_{T}}{2T}\frac{\partial\tau_{0}}{\partial r}\frac{\delta u_{\phi4(1)}}{r}\left(I_{3}'-I_{5}'\right)\nonumber\\&&-4\pi p_{T}K_{0}\left(\frac{m_{T}}{T}\right)\int dr\  r\tau_{0}\frac{p_{T}}{2T}\frac{\partial\tau_{0}}{\partial r}\delta u_{r4(1)}\left(I_{3}'+I_{5}'\right)\nonumber\\&&-4\pi p_{T}K_{0}\left(\frac{m_{T}}{T}\right)\int dr\  r\tau_{0}\left(\frac{\partial\delta\tau_{4(1)}}{\partial r}+\frac{\partial\tau_{0}}{\partial r}\frac{\delta\tau_{4(1)}}{\tau_{0}}\right)I_{4}'.\nonumber\\
    \delta J_{2}^{(4)*}&=&-4\pi p_{T}K_{0}(\frac{m_{T}}{T})\int dr\  r\tau_{0}\left(\frac{\partial\delta\tau_{4(2)}}{\partial r}+\frac{\partial\tau_{0}}{\partial r}\frac{\delta\tau_{4(2)}}{\tau_{0}}\right)I_{4}'\nonumber\\&&-4\pi p_{T}K_{0}(\frac{m_{T}}{T})\int dr\  r\tau_{0}\frac{p_{T}}{2T}\frac{\partial\tau_{0}}{\partial r}\left(\delta u_{r4(2)}\left(I_{3}'+I_{5}'\right)-\frac{\delta u_{\phi4(2)}}{r}\left(I_{3}'-I_{5}'\right)\right)\nonumber\\&&-4\pi p_{T}K_{0}(\frac{m_{T}}{T})\int dr\  r\tau_{0}\frac{p_{T}}{2T}\left(\frac{1}{2}\frac{\partial\delta\tau_{2}}{\partial r}+\frac{1}{2}\frac{\delta\tau_{2}}{\tau_{0}}\frac{\partial\tau_{0}}{\partial r}\right)\left(\delta u_{r2}\left(I_{3}'+I_{5}'\right)-\frac{\delta u_{\phi2}}{r}\left(I_{3}'-I_{5}'\right)\right)\nonumber\\&&-4\pi p_{T}K_{0}(\frac{m_{T}}{T})\int dr\  r\tau_{0}\frac{1}{2}\frac{\delta\tau_{2}}{\tau_{0}}\frac{\partial\delta\tau_{2}}{\partial r}I_{4}'\nonumber\\&&-4\pi p_{T}K_{0}\left(\frac{m_{T}}{T}\right)\int_{\Sigma}dr\ r\tau_{0}\left(\frac{p_{T}}{2T}\right)^{2}\frac{1}{4}\frac{\partial\tau_{0}}{\partial r}\left(\delta u_{r2}^{2}+(\frac{\delta u_{\phi2}}{r})^{2}\right)\left(I_{2}'+I_{6}'\right)\nonumber\\&&-4\pi p_{T}K_{0}\left(\frac{m_{T}}{T}\right)\int_{\Sigma}dr\ r\tau_{0}\left(\frac{p_{T}}{2T}\right)^{2}\frac{1}{2}\frac{\partial\tau_{0}}{\partial r}\left[\left(\delta u_{r2}^{2}-(\frac{\delta u_{\phi2}}{r})^{2}\right)I_{4}'-\delta u_{r2}\frac{\delta u_{\phi2}}{r}\left(I_{2}'-I_{6}'\right)\right].
\end{eqnarray}
For $J_3$, we have, 
\begin{eqnarray}
    \delta J_{3}^{(2)}&=&-4\pi p_{T}K_{0}\left(\frac{m_{T}}{T}\right)\int dr\ \tau_{0}\delta\tau_{2}\frac{4}{z}I_{2},\nonumber\\
    \delta J_{3}^{(4)}&=&-4\pi p_{T}K_{0}\left(\frac{m_{T}}{T}\right)\int dr\ \tau_{0}\delta\tau_{4(1)}\frac{16}{z}I_{4},\nonumber\\
    \delta J_{3}^{(4)*}&=&-4\pi p_{T}K_{0}\left(\frac{m_{T}}{T}\right)\int dr\ \tau_{0}\frac{p_{T}}{2T}\delta\tau_{2}\left(\delta u_{r2}\left(I_{3}'-I_{5}'\right)-\frac{\delta u_{\phi2}}{r}\left(I_{3}'+I_{5}'-2I_{4}\right)\right)\nonumber\\&&-4\pi p_{T}K_{0}\left(\frac{m_{T}}{T}\right)\int dr\ \tau_{0}\left(\delta\tau_{4(2)}\frac{16}{z}+\frac{(\delta\tau_{2})^{2}}{\tau_{0}}\frac{4}{z}\right)I_{4}.
\end{eqnarray}

\section{Derive $\chi_{ij}$ in small and large $p_T$ limits} \label{sec:appendix_limits}
In the small-$p_T$ limit, the modified Bessel function of the first kind $I_{n}(z)$ can be approximated as $I_{n}(z)\approx\frac{1}{n!}(\frac{z}{2})^{n}$, where $z=p_{T}u_{r0}/T\sim p_{T}/(B^{3}T)$. Given our early-time assumption $B\gg1$, we have $z\ll1$ for low $p_T$. Under these conditions, we expand all $I_{n}$ functions and carry out the $r$ integration to obtain simplified expressions for $J^{i}$,
\begin{eqnarray}
\delta J_{1}^{(2)}&\approx&-4\pi m_{T}K_{1}\left(\frac{m_{T}}{T}\right)\frac{2^{15}L^{3}p_{T}^{2}}{21B^{9}T^{2}},\nonumber\\
\delta J_{1}^{(4)}&\approx&-4\pi m_{T}K_{1}\left(\frac{m_{T}}{T}\right)\frac{2^{23}\times21L^{3}p_{T}^{4}}{7735B^{15}T^{4}},\nonumber\\
\delta J_{1}^{(4)*}&\approx&4\pi m_{T}K_{1}\left(\frac{m_{T}}{T}\right)\frac{2^{22}\times113L^{3}p_{T}^{4}}{7735B^{15}T^{4}},\nonumber\\J_{1}^{(0)}&\approx&4\pi m_{T}K_{1}\left(\frac{m_{T}}{T}\right)\frac{16L^{3}}{B^{3}},\nonumber\\
\delta J_{2}^{(2)}&\approx&-4\pi K_{0}\left(\frac{m_{T}}{T}\right)\frac{2^{13}L^{3}p_{T}^{2}}{3B^{9}T},\nonumber\\
\delta J_{2}^{(4)}&\approx&-4\pi K_{0}\left(\frac{m_{T}}{T}\right)\frac{2^{22}\times268L^{3}p_{T}^{4}}{7735B^{15}T^{3}}.\nonumber\\
\delta J_{2}^{(4)*}&\approx&4\pi K_{0}\left(\frac{m_{T}}{T}\right)\frac{2^{22}\times505L^{3}p_{T}^{4}}{7735B^{15}T^{3}}.\nonumber\\J_{2}^{(0)}&\approx&4\pi K_{0}\left(\frac{m_{T}}{T}\right)\frac{2^{15}L^{3}p_{T}^{2}}{21B^{9}T},\nonumber\\
\delta J_{3}^{(4)}&\approx&-4\pi K_{0}\left(\frac{m_{T}}{T}\right)\frac{2^{22}\times4L^{3}p_{T}^{4}}{455B^{15}T^{3}},\nonumber\\
\delta J_{3}^{(4)*}&\approx&4\pi K_{0}\left(\frac{m_{T}}{T}\right)\frac{2^{22}\times5L^{3}p_{T}^{4}}{91B^{15}T^{3}},\nonumber\\
\delta J_{3}^{(2)}&\approx&-4\pi K_{0}\left(\frac{m_{T}}{T}\right)\frac{2^{12}\times6L^{3}p_{T}^{2}}{7B^{9}T}a_{2}.\label{eq:J_small_pT}
\end{eqnarray}

In the high-$p_{T}$ limit, the modified Bessel function of the first kind can be approximated as $I_{n}(z)\approx\frac{e^{z}}{\sqrt{2\pi z}}$, which is independent of the order $n$ at leading exponential accuracy. Consequently, the expressions for $J^{i}$ simplify to the following $r$ integrals,
\begin{eqnarray}
J_{1}^{(0)}&\approx&4\pi m_{T}K_{1}\left(\frac{m_{T}}{T}\right)\int dr\ r\tau_{0}\frac{e^{z}}{\sqrt{2\pi z}}+\mathcal{O}(a_{2}^{2}),\nonumber\\
\delta J_{1}^{(2)}&\approx&4\pi m_{T}K_{1}\left(\frac{m_{T}}{T}\right)\int dr\ r\tau_{0}\frac{p_{T}}{2T}\ 2\delta u_{r2}\frac{e^{z}}{\sqrt{2\pi z}},\nonumber\\
\delta J_{1}^{(4)}&\approx&4\pi m_{T}K_{1}\left(\frac{m_{T}}{T}\right)\int dr\ r\tau_{0}\frac{p_{T}}{2T}\ 2\delta u_{r4(1)}\frac{e^{z}}{\sqrt{2\pi z}},
\end{eqnarray}
and,
\begin{eqnarray}
\delta J_{1}^{(4)*}&\approx&4\pi m_{T}K_{1}\left(\frac{m_{T}}{T}\right)\int dr\ r\tau_{0}\left(\frac{p_{T}}{2T}\right)^{2}\ \delta u_{r2}^{2}\frac{e^{z}}{\sqrt{2\pi z}},\nonumber\\
J_{2}^{(0)}&\approx&-4\pi p_{T}K_{0}\left(\frac{m_{T}}{T}\right)\int dr\ r\tau_{0}\frac{\partial\tau_{0}}{\partial r}\frac{e^{z}}{\sqrt{2\pi z}}+\mathcal{O}(a_{2}^{2}),\nonumber\\
\delta J_{2}^{(2)}&\approx&-4\pi p_{T}K_{0}\left(\frac{m_{T}}{T}\right)\int dr\ r\tau_{0}2\frac{p_{T}}{2T}\frac{\partial\tau_{0}}{\partial r}\delta u_{r2}\frac{e^{z}}{\sqrt{2\pi z}},\nonumber\\
\delta J_{2}^{(4)}&\approx&-4\pi p_{T}K_{0}\left(\frac{m_{T}}{T}\right)\int dr\ 2r\tau_{0}\frac{p_{T}}{2T}\frac{\partial\tau_{0}}{\partial r}\delta u_{r4(1)}\frac{e^{z}}{\sqrt{2\pi z}}\nonumber\\
\delta J_{2}^{(4)*}&\approx&-4\pi p_{T}K_{0}\left(\frac{m_{T}}{T}\right)\int_{\Sigma}dr\ r\tau_{0}\left(\frac{p_{T}}{2T}\right)^{2}\frac{\partial\tau_{0}}{\partial r}\left(\delta u_{r2}^{2}\right)\frac{e^{z}}{\sqrt{2\pi z}},\nonumber\\
\delta J_{3}^{(2)}&\approx&-4\pi p_{T}K_{0}\left(\frac{m_{T}}{T}\right)\int dr\ \tau_{0}\delta\tau_{2}\frac{4}{z}\frac{e^{z}}{\sqrt{2\pi z}},\nonumber\\
\delta J_{3}^{(4)}&\approx&-4\pi p_{T}K_{0}\left(\frac{m_{T}}{T}\right)\int dr\ \tau_{0}\delta\tau_{4(1)}\frac{16}{z}\frac{e^{z}}{\sqrt{2\pi z}},\nonumber\\
\delta J_{3}^{(4)*}&\approx&-4\pi p_{T}K_{0}\left(\frac{m_{T}}{T}\right)\int dr\ \tau_{0}\frac{4}{z}\left(4\delta\tau_{4(2)}+\frac{(\delta\tau_{2})^{2}}{\tau_{0}}\right)\frac{e^{z}}{\sqrt{2\pi z}}.
\end{eqnarray}
We find that these integrals diverge. To handle these divergences, we evaluate the integral using the saddle-point approximation,
\begin{eqnarray}
\int g(x)e^{f(x)}dx & \approx & g(x_{0})e^{f(x_{0})}\sqrt{\frac{2\pi}{f^{\prime\prime}(x_{0})}},\label{eq:Saddle_point}
\end{eqnarray}
where $g(x)$ and $f(x)$ are smooth functions and $x_{0}$ is the saddle point satisfying $f^\prime(x_{0})=0$. This approximation reduces the integral to the local behavior of $f(x)$ and $g(x)$ near $x_0$.
The saddle point can be obtained by solving the following equation, 
\begin{eqnarray}
\frac{d}{dr}z^{*} & = & \frac{d}{dr}\left[\frac{p_{T}}{T}\frac{2r(2L)^{5}}{B^{3}(L^{2}+r^{2})^{3}}\right]=0.
\end{eqnarray}
where we get the saddle point to be $r^{*}=L/\sqrt{5}$. 

With
this method, we could easily get, $J_{2}^{(0)}/J_{1}^{(0)}\sim\frac{\partial\tau_{0}^{*}}{\partial r}\sim B^{-3}\ll1$,
$\delta J_{2}^{(2)}/\delta J_{1}^{(2)}\sim B^{-3}\ll1$ and $\delta J_{3}^{(2)}/\delta J_{1}^{(2)}\sim B^{-3}\ll1$,
thus from the Eqs.~(\ref{eq:v2_v4_01}), we 
get the result for $v_{2}$ in large $p_{T}$ limit,
\begin{eqnarray}
v_{2} & = & \frac{a_{2}}{2}\frac{\delta J_{1}^{(2)}}{J_{1}^{(0)}}\approx\frac{p_{T}a_{2}}{2T}\delta u_{r2}^{*}=\epsilon_{2}\frac{p_{T}}{T}\frac{500\sqrt{5}}{B^{3}81}.
\end{eqnarray}
Likely, while $n=4$, we also make the integral using the saddle point
$r^{*}=L/\sqrt{5}$ and get the result, 
\begin{eqnarray}
v_{4}&\approx&\frac{1}{2}\frac{\delta J_{1}^{(4)*}}{J_{1}^{(0)}}a_{2}^{2}\approx\frac{1}{2}\left(\frac{p_{T}}{2T}\right)^{2}\delta u_{r2}^{*2}a_{2}^{2}=\frac{1}{2}v_{2}^{2}.
\end{eqnarray}

\bibliographystyle{h-physrev}
\bibliography{ref}

\end{document}